\documentstyle[prl,aps,preprint]{revtex}

\begin{document}
%\draft
\title{Can We Distinguish Between the Grand Canonical \\ and
the Canonical Ensemble in a BEC Experiment?
}
\author
{
C. Herzog and M. Olshanii
}
\address{
{\small Lyman Laboratory, Harvard University,}
{\small Cambridge MA 02138, USA} \\
maxim@atomsun.harvard.edu
}
\date{\today}
\maketitle 
%%%%%%%%%%%%%%%%%%%%%%%%%%%%%%%%%%%%%%%%%%%%%%%%%%%%%%%%%%%%%%%%%%%%%%%%%%%
\begin{abstract}
For ensemble of bosons
trapped in a 1D harmonic potential well
we have found an analytical formula for the canonical partition function
and shown that, for 100 trapped atoms, the
discrepancy between
the grand canonical and the canonical predictions for the condensate
fraction reaches 10\%
in the vicinity of the Bose-Einstein threshold.
This discrepancy decreases only logarithmically as the
number of atoms increases. Furthermore we
investigate numerically the case of a 3D ``cigar-shape'' trap in the
range of parameters corresponding to current BEC experiments.
\end{abstract}
%%%%%%%%%%%%%%%%%%%%%%%%%%%%%%%%%%%%%%%%%%%%%%%%%%%%%%%%%%%%%%%%%%%%%%%%%%

%%%%%%%%%%%%%%%%%%%%%%%%%%%%%%%%%%%%%%%%%%%%%%%%%%%%%%%%%%%%%%%%%%%%%%%%%
Recently, Bose-Einstein condensation (BEC) in trapped atomic gases
\cite{ERIC,WOLF} has been realized.  The trapped atomic cloud possesses
two remarkable features:  First, the system is small enough so that finite
particle effects are potentially observable, and second, particle
interactions are weak.  The thermodynamics of such a system is an
interesting and rich area for scientific analysis. 

The equivalence of the grand canonical and fixed-$N$ canonical
descriptions of a statistical system is an old question widely discussed
in the textbooks on statistical mechanics \cite{KITTEL,HUANG}.  For a
bosonic gas, where grand canonical fluctuations of the 
ground state population
become large at and below the Bose-Einstein threshold, such an
equivalence is not obvious. 
It is shown that in the thermodynamic limit
$N \rightarrow \infty$ both ensembles give the same predictions 
for the mean values of occupation numbers even in the absence  
of particle interactions \cite{FUJI}.
Furthermore it is well-known 
that for large $N$, 
interactions between particles
lead to suppression of fluctuations in the grand canonical ensemble 
\cite{HUANG}. However, for a
finite system with a mesoscopic number of particles, 
the equivalence of the two ensembles is not ensured. 

The main scaling laws for the fluctuations in an ideal canonical bose gas
are derived by Fujiwara {\it et al.} \cite{FUJI}.  
The fixed-$N$ bose statistics is shown to be
closely related to Gentile's grand canonical intermediate statistics
\cite{GENT}.  Krauth has performed  fixed-$N$ finite 
temperature Monte-Carlo calculations for a
3D harmonic potential \cite{KRAUTH}.  Although the main
subject of the paper \cite{KRAUTH} is the role of interactions, 
it is shown also that
for macroscopic numbers of particles, the noninteracting grand canonical and
canonical ensembles agree very well. These conclusions are consistent with
the numerical results of Politzer \cite{POLITZER}. In the present paper we
consider {\it mesoscopic} values of number of particles ($N\sim 100$) 
confined in
a one-dimensional harmonic trap \cite{1D} and in a three-dimensional
``cigar-shape'' trap. We show that the 
grand canonical/canonical deviations in this case are {\it substantial}.

%%%%%%%%%%%%%%%%%%%%%%%%%%%%%%%%%%%%%%%%%%%%%%%%%%%%%%%%%%%%%%%%%%%%%%%%%%
Consider an ensemble of $N$ noninteracting bosons confined in
a 1D harmonic potential in thermal (but not in diffusive)
equilibrium with a large reservoir.  The population distribution among the
different energy levels of the $N$-particle system will be
given by the Boltzmann law:
\begin{eqnarray}
\rho([n] ) \sim \exp \left\{ -\beta E([n]) \right\} \\
\left[ n \right] = \{ n_0, n_1, ..., n_s, ... | \sum_s n_s = N\}  \, ,
\end{eqnarray}
where $E([n]) = \sum_s n_s \epsilon_s$ is the $N$-particle energy for 
the given configuration of occupation numbers $\left[ n \right]$,
$\epsilon_s = \hbar \omega s \, (s = 0, 1, 2, ...)$ 
is the single particle energy spectrum, 
$\omega$ is the harmonic oscillator frequency, $\beta = 1/k_B T$, and $T$ is
the temperature of the system.  Note that in the harmonic oscillator case,
the $N$-particle energy is quantized as 
\begin{eqnarray}
E([n]) = \hbar \omega K \, ,
\end{eqnarray}
where $K = \sum_s n_s s$.

To calculate mean occupation numbers 
of the oscillator states, we
need to know the partition function $Q$ and its derivatives.  We show below 
that in the 1D harmonic oscillator case, occupation numbers may be
calculated analytically as finite sums of finite products. 
The canonical partition function $Q(\beta, N)$ can be represented by a power 
series of $x= \exp(-\beta \hbar \omega)$:
\begin{eqnarray}
Q(\beta, N) = \sum_{K=0}^{\infty} x^K \,
\Gamma(K, N) \, .
\label{canon}
\end{eqnarray}
The microcanonical partition function 
\begin{eqnarray}
\Gamma(K, N) &=& \sum_{
\stackrel{[n]} {\Sigma n_s = N, \Sigma n_s s = K}
} 1 \nonumber \\
&=& \sum_{N^{\prime}=0}^{N} \sum_{
\stackrel{[n]^{\prime}} 
{\Sigma^{\prime} n_s = N^{\prime}, \Sigma^{\prime} n_s s = K} 
} 1
\label{microcanon}
\end{eqnarray}
equals the number of representations (partitions) of $K$ as 
an unordered sum
of at most $N$ positive integers. 
Here $N^{\prime}$ is the total population of the excited 
states, $\left[ n \right]^{\prime} = 
\{ n_1, n_2, ..., n_s, ...\}$
is a particular configuration of excited state occupation numbers,
and the ``primed'' sum $\Sigma^{\prime} = \Sigma_{s=1}^{\infty}$   
denotes a sum over the excited states.

According to a well-known number theory theorem \cite{NUMBERS} 
{\it the number $\Gamma(K, N)$ of partitions of $K$ with at most $N$
parts equals the number ${\cal P}(K, N)$ of partitions of $K$ with 
parts not exceeding $N$}.  Hence, the canonical partition function
(\ref{canon}) is nothing else but the generating function for the 
restricted partition function ${\cal P}(K, N)$ \cite{NUMBERS}:
\begin{eqnarray}
Q(\beta, N) &=& \sum_{K=0}^{\infty} x^K \, {\cal P}(K, N)
\nonumber \\
&=& \prod_{\tilde{N} = 1}^{N} 
\frac{1}{1-x^{\tilde{N}}}  \, \, .
\label{QQ}
\end{eqnarray}

Derivatives of the partition function 
$Q_s = 
-\beta^{-1} (\partial Q / \partial \epsilon_s)  $
can not be found directly from the expression (\ref{QQ})
which is specific for the 1D harmonic oscillator. 
Instead, we have found a {\it general} recursion relation between
the canonical partition function and its derivatives:  
\begin{eqnarray}
Q_s(\beta, N+1) = \exp(-\beta \epsilon_s) 
( Q_s(\beta, N) + Q(\beta, N) ) \, .
\end{eqnarray}
This relation can be 
applied to any fixed-$N$, noninteracting, bosonic system.

Finally, the mean occupation numbers are given by
\begin{eqnarray}
\langle n_s \rangle &=& \frac{Q_s}{Q} \nonumber \\
&=& \sum_{\tilde{N} = 1}^{N} 
x^{(N \! -\! \tilde{N} \! + \!1)s}
\prod_{\tilde{\tilde{N}}=\tilde{N}}^{N} 
(1-x^{\tilde{\tilde{N}}}) \, .
\end{eqnarray}
This expression is easy to analyze in the continuous limit 
with respect to $N$.  For example, below the BEC threshold, 
the condensate population is approximately given by
\begin{eqnarray}
\frac{\langle n_0 \rangle}{N} &\approx&  
1 - \frac{k_B T}{N \hbar \omega} 
\left( \log( k_B T/\hbar \omega) + C + o(1) \right) 
\nonumber \\
&\stackrel{N \rightarrow \infty}{\longrightarrow}&
 1 - \frac{T}{T_c} \, , 
\label{THERM_LIM1}
\end{eqnarray}
where $C \approx 0.5772$ is the Euler constant.  The transition 
temperature is given by
\begin{eqnarray}
N = \frac{k_B T_c}{\hbar \omega}  
\log \left( \frac{ const \, k_B T_c}{ \hbar \omega } \right) \, ,
\label{T_c}
\end{eqnarray}
where the choice of $const$ is a matter of convention. Note that
the thermodynamic limit (\ref{THERM_LIM1}) coincides with 
the one predicted for grand canonical statistics \cite{1D}.

Now we are ready to compare the canonical and grand canonical predictions for
the condensate population $\langle n_0 \rangle$.  In Fig.~1 we plot 
the population 
of the ground state for different numbers of particles.  For the
grand canonical predictions we simply repeat the finite-system
calculations of \cite{1D}.  To facilitate the comparison, 
we made the same choice $const = 2$ in expression (\ref{T_c}). 
Both curves approach the 
thermodynamic limit (\ref{THERM_LIM1}) 
as the number of particles increases.
However, for a finite number of particles 
the discrepancy between the two models is 
quite significant.  In the vicinity of the BEC threshold, the 
relative deviation 
$(\langle n_0^{gr. \, canon.} \rangle  - \langle n_0^{canon.} \rangle)
/ \langle n_0^{canon.} \rangle$ decreases slowly with $N$ and goes
from $10\%$ for $100$ atoms to $5\%$ for $10,000$ atoms. 
We have checked
that this deviation decreases according to a $1/\ln(N)$ scaling law
for a fixed  $T/T_c$.  Note that the rate at which both the grand 
canonical \cite{1D} and canonical (\ref{THERM_LIM1}) populations approach 
the thermodynamic limit also obeys this law.

%%%%%%%%%%%%%%%%%%%%%%%%%%%%%%%%%%%%%%%%%%%%%%%%%%%%%%%%%%%%%%%%%%%%%%%
We turn now to the 3D trap.  To our knowledge there is no 
simple analytic expression for the canonical partition
function in this case.  Numerically, it can be calculated
by integration of the grand canonical partition function in
the complex plain of chemical potential \cite{FUJI,POLITZER}.  Indeed
\begin{eqnarray}
Q(\beta, N) 
&=& \sum_{
\stackrel{[n]} {\Sigma n_s = N}} \exp \left\{ -\beta E([n]) \right\}
\nonumber \\
&=& \sum_{[n]}
\delta_{\Sigma n_s, \, N} \, \exp \left\{ -\beta E([n]) \right\}
\nonumber \\
&=& \frac{\beta}{2\pi i} \int_{-\pi i}^{+\pi i}
d\mu \, \exp(-N\mu) Z(\beta, \mu) \, ,
\end{eqnarray}
where 
\begin{eqnarray}
Z(\beta, \mu) = \prod_{s_x,s_y,s_z=0}^{\infty}
\frac{1}
{1 - \exp[-\beta (\sum_{\alpha=x,y,z} \hbar \omega_{\alpha} s_{\alpha} -\mu)]}
\end{eqnarray}
is the grand canonical partition function,
$\omega_{\alpha}\,(\alpha=x,y,z$) are the trap frequencies, and 
the expression 
$\delta_{q,q^{\prime}} = 
(2\pi i)^{-1} \int_{-\pi i}^{+\pi i} d\xi \exp[(q-q^{\prime})\xi]$ 
for the the Kronecker
delta has been used.  
Derivatives of the partition function
can be expressed through $Z(\beta, \mu)$ in the same way.

In Fig. 2 we plot the condensate fraction as a function of temperature for
both grand canonical and canonical ensembles.  We have chosen the
``cigar-shape'' configuration $\omega_{\perp} = 17.78 \, \omega_z$, where
$\omega_{\perp} = \omega_x = \omega_y$.  The three dimensional
Bose-Einstein transition temperature is given by 
\begin{eqnarray} 
N &=& g_3(1) \frac{(k_B T_c)^3}{\hbar^3 \prod_{\alpha=x,y,z} \omega_{\alpha}}
\nonumber \\ 
&+& \frac{g_2(1)}{2} 
\frac{(k_B T_c)^2 \sum_{\alpha=x,y,z} \omega_{\alpha}} 
{\hbar^3 \prod_{\alpha=x,y,z} \omega_{\alpha}} 
+ {\cal O}(k_B T_c/\hbar \omega) \, ,
\end{eqnarray} 
where the second line is the
finite-$N$ correction \cite{1D}. 
Here $g_d(z) = \sum_{j=1}^{\infty} z^j/j^d$ is
the Bose-Einstein function. For comparison, we have also plotted the
thermodynamic limit 
\begin{eqnarray} \frac{\langle n_0 \rangle}{N} = 
1 - \left( \frac{T}{T_c} \right)^3  \, .
\label{THERM_LIM2} 
\end{eqnarray} 
For $100$ particles, 
depending on the temperature, the system exhibits
both 3D and 1D characteristics.  At $T \sim 0.4 \, T_c$ the temperature
reaches the zero-point energy $\hbar \omega_{\perp} /2$ for tranverse
oscillations. 
The grand canonical/canonical discrepancy is less than in the
purely 1D system but is still close to 10\%. 

In the above discussion we neglected  
particle interactions.  To estimate the importance of interactions in our 
model, we consider the ``worst'' case of 
zero temperature where the spatial density is the highest and therefore 
the interactions are strongest.  For typical  
Ioffe-Pritchard trap parameters \cite{M_TRAPS} 
($\omega_{\perp} = 2\pi \times 101 \,  Hz, \, 
\omega_z = 2\pi \times 5.7 \, Hz$, $N=100$) for sodium atoms (scattering length
$a = 92 \, Bohr$, atomic mass $M = 23 \, amu$) the mean-field
corrections to the oscillation frequencies are
quite small:
$\delta \omega_{\perp} = 0.03 \, \omega_{\perp}$ and 
$\delta \omega_z = 0.15 \, \omega_{\perp}$.  To estimate the corrections, 
we minimized the Gross-Pitaevskii 
energy functional with a ground state oscillator 
wave function seeded with unknown frequencies \cite{BAYM}. 
Note that for the parameters chosen, the system exhibits
a BEC transition at $T_c = 6.5 \, nK$.

%%%%%%%%%%%%%%%%%%%%%%%%%%%%%%%%%%%%%%%%%%%%%%%%%%%%%%%%%%%%%%%%%%%%%%%
We acknowledge fruitful discussions with 
H.D.~Politzer, T.H.~Bergeman, J.H.~Thywissen, E.~Heller, L.~You, 
M.~Prentiss and W.~Ketterle. 
M.O. was supported by the National Science Foundation
grant for light force dynamics \#PHY-93-12572. C.H. was
supported by Harvard University.
This work was partially supported by
the NSF through
a grant for the Institute for Theoretical Atomic and Molecular
Physics at Harvard University and the Smithsonian Astrophysical Observatory.
%%%%%%%%%%%%%%%%%%%%%%%%%%%%%%%%%%%%%%%%%%%%%%%%%%%%%%%%%%%%%%%%%%%%%%%%%%
\begin{figure}
\caption{The condensate fraction for the 1D harmonic
oscillator as a function of  temperature.  Both grand canonical and 
canonical predictions are
shown.  The straight line is the thermodynamic limit 
(9).
}
\caption{The condensate fraction for a 3D ``cigar-shape'' trap
as a function of temperature.
Both grand canonical and
canonical predictions are
shown.
Here 
$\omega_{\perp} = 2\pi \times 101 \,  Hz, \,
\omega_z = 2\pi \times 5.7 \, Hz, \, N=100, \, T_c = 6.5 \, nK$.
The thermodynamic limit $N=\infty$
(14) is also shown.
} 
\end{figure}
%
%%%%%%%%%%%%%%%%%%%%%%%%%%%%%%%%%%%%%%%%%%%%%%%%%%%%%%%%%%%%%%%%%%%%%%%%%%%%
%

%
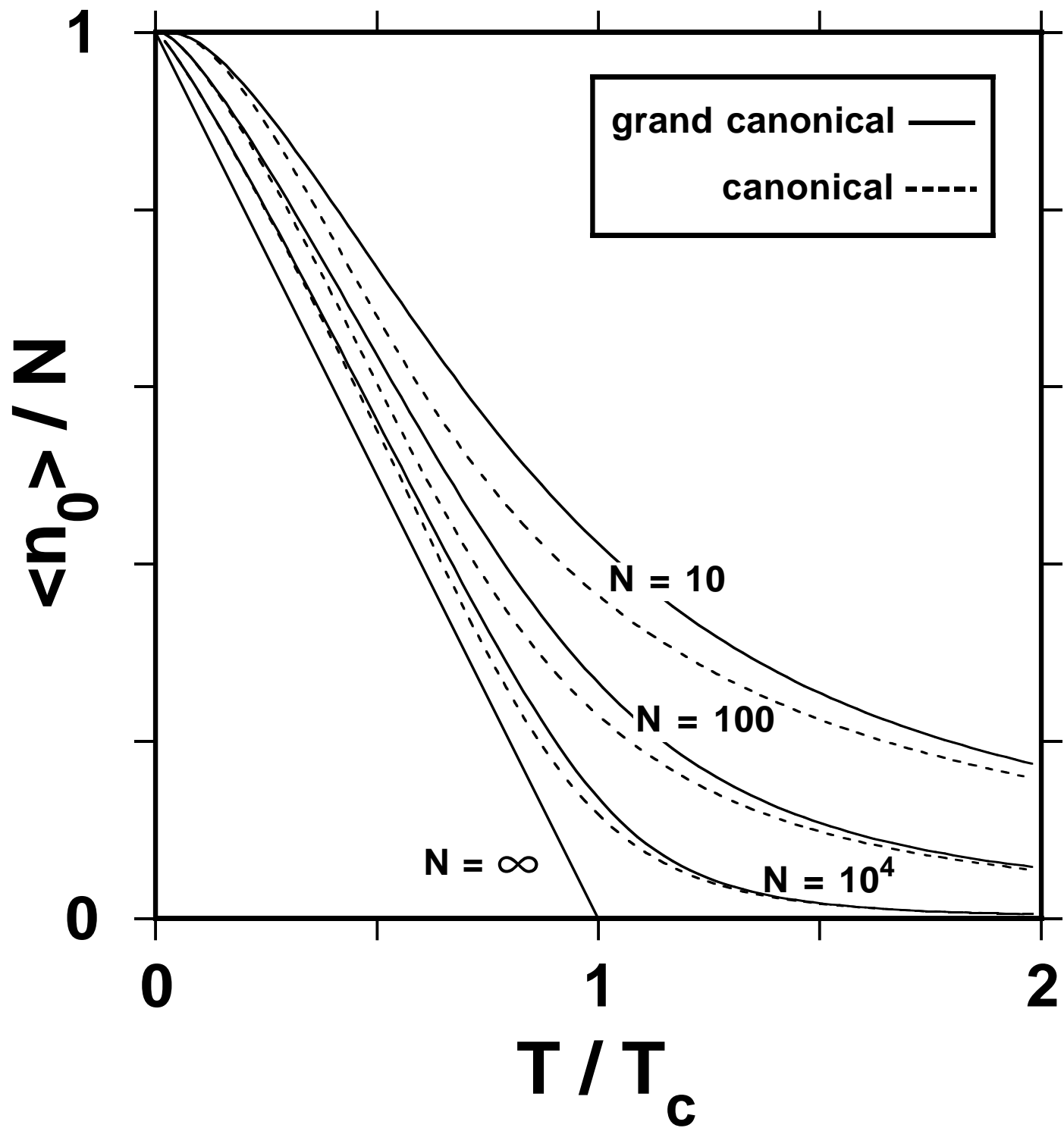
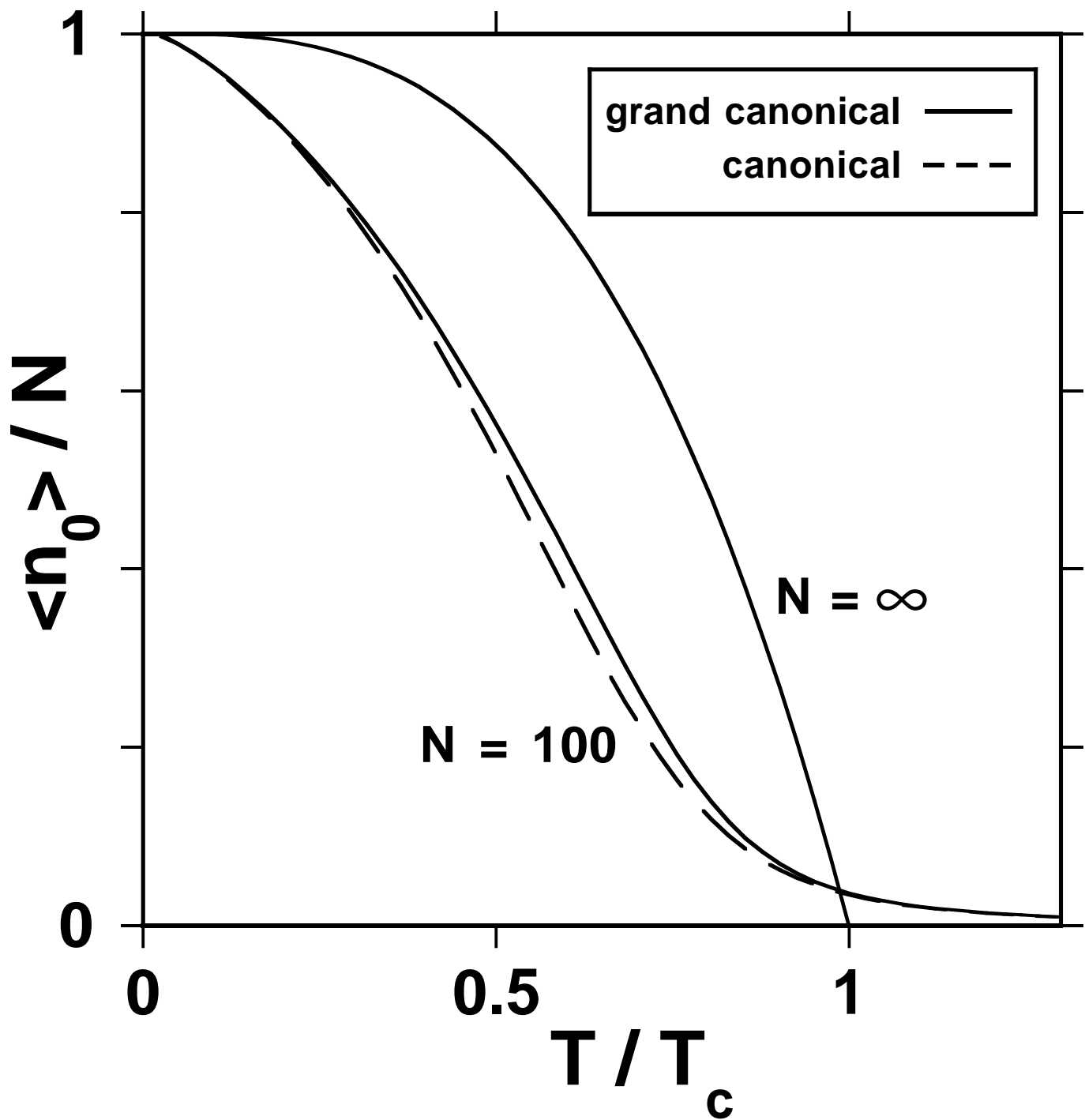
\end{document}